\documentclass[conference]{IEEEtran}
\IEEEoverridecommandlockouts
\usepackage{cite}
\usepackage[numbers,sort&compress]{natbib}
\usepackage{amsmath,amssymb,amsfonts}
\usepackage{algorithmic}
\usepackage{graphicx}
\usepackage{multirow}
\usepackage{graphicx}
\usepackage{textcomp}
\usepackage{amsmath,graphicx}
\usepackage{xcolor}
\usepackage{enumitem,amssymb,cite}
\usepackage{multirow}
\usepackage{bigstrut}
\usepackage{bigdelim}
\usepackage{booktabs,chemformula}
\usepackage{tikz}
\usepackage{hhline}
\usepackage{diagbox}
\usepackage{graphicx}
\usepackage{soul}
\usepackage{caption}
\def\BibTeX{{\rm B\kern-.05em{\sc i\kern-.025em b}\kern-.08em
    T\kern-.1667em\lower.7ex\hbox{E}\kern-.125emX}}
\newcommand{\bs}{\boldsymbol} 

\begin{document}
\title{Synergistic Effects of Knowledge Distillation and Structured Pruning for Self-Supervised Speech Models\\
% {\footnotesize \textsuperscript{*}Note: Sub-titles are not captured for https://ieeexplore.ieee.org  and
}
\author{\IEEEauthorblockN{Shiva Kumar C}
\IEEEauthorblockA{\textit{Samsung Research} \\
Bangalore, India \\
c.shiva1393@gmail.com}
\and
\IEEEauthorblockN{Jitendra Kumar Dhiman}
\IEEEauthorblockA{\textit{Samsung Research} \\
Bangalore, India \\
jkdiith@gmail.com}
\and
\IEEEauthorblockN{Nagaraj Adiga}
\IEEEauthorblockA{\textit{Krutrim} \\
Bangalore, India \\
nvadigauvce@gmail.com}
%\and
%\IEEEauthorblockN{Jainag Ambati}
%\IEEEauthorblockA{\textit{Samsung Research} \\
%Bangalore, India \\
%a.jainag@samsung.com}
\and
\IEEEauthorblockN{Shatrughan Singh}
\IEEEauthorblockA{\textit{Samsung Research} \\
Bangalore, India \\
shatrughan.s@samsung.com}

}
\maketitle
\begin{abstract}
Traditionally, \textbf{K}nowledge \textbf{D}istillation (KD) is used for model compression, often leading to suboptimal performance. 
In this paper, we evaluate the impact of combining KD loss with alternative pruning techniques, including Low-Rank Factorization (LRF) and $l_0$ regularization, on a conformer-based pre-trained network under the paradigm of Self-Supervised Learning (SSL).
We also propose a strategy to jointly prune and train an RNN-T-based ASR model, demonstrating that this approach yields superior performance compared to pruning a pre-trained network first and then using it for ASR training.
This approach led to a significant reduction in word error rate: 
$l_0$ and KD combination achieves the best non-streaming performance, with a $8.9\%$ \textbf{R}elative \textbf{W}ord \textbf{E}rror \textbf{R}ate (RWER) improvement over the baseline, while LRF and KD combination yields the best results for
streaming ASR, improving RWER by $13.4\%$.
\end{abstract}
\begin{IEEEkeywords}
low-rank factorization, knowledge distillation, self-supervised learning, structured pruning, $l_0$ regularization
\end{IEEEkeywords}
\section{Introduction}
\label{sec:intro}
\textbf{S}elf-\textbf{S}upervised \textbf{L}earning (SSL) has proven to be an effective approach for leveraging vast amounts of unlabeled data to train deep neural networks, which are subsequently fine-tuned for specific downstream tasks~\cite{wav2vec2, hubert, wavlm, data2vec}. 
This paradigm often results in models with billions of parameters, leading to bigger models. 
Such extensively large models present significant challenges for on-device deployment, primarily due to memory and latency constraints.
As a result, there is a critical need to compress these models effectively, ensuring that performance remains largely unaffected while addressing the deployment challenges.
\par 
A widely adopted method for model compression is \textbf{K}nowledge \textbf{D}istillation (KD), where the KD loss is computed as the Kullback-Leibler divergence between the teacher and student model outputs~\cite{distilhubert, hinton2015distilling, adiga2023compression, FitHuBERT,rathod2022multi}.
While effectively reducing the model size, this approach typically requires manual design of the student model's architecture, where specific layers and parameters are chosen based on intuition or trial and error~\cite{rathod2022multi}. 
This manual intervention often results in suboptimal performance of the model, emphasizing the need to explore alternative compression methods.\par 
% Moreover, KD tends to perform poorly at high compression rates (typically above $50\%$)~\cite{rathod2022multi} 
To address the limitations of KD, pruning was introduced~\cite{LRF_structured, L0_reg_louizos2017learning,lai2021parp,wang2024unstructured}.
In particular, the method of structured pruning automates the pruning of entire network structures, such as neurons, channels, or layers, reducing the need for manual architecture tuning.
A recently proposed technique utilizes \textbf{L}ow-\textbf{R}ank \textbf{F}actorization (LRF) of model weights to achieve structured pruning~\cite{LRF_structured}.
To further enhance the effectiveness of pruning techniques, the primary loss function is often augmented with an $l_0$ regularization to promote sparsity in the network architecture, achieving higher compression rates ~\cite{dphubert}.
Notable works exploring these techniques include~\cite{DeepversusWide, onceforall, LightHuBERT, dphubert}, primarily applied to a pre-trained network to maintain task-agnostic capabilities. 
However, the impact of model compression during the fine-tuning stage has been less extensively studied, 
particularly when the downstream task is predetermined such as the \textbf{A}utomatic \textbf{S}peech \textbf{R}ecognition (ASR).\\
This paper extends the existing research by making the following contributions.
% \begin{enumerate}[itemsep=2pt,parsep=0.01pt,partopsep=0.01pt,leftmargin=*,topsep=0pt]
\begin{enumerate}
    \item We analyze the impact of combining KD loss with either LRF- or $l_0$ regularization-based pruning of a pre-trained network, demonstrating that this combination leads to superior performance for ASR over using either LRF or $l_0$ regularization (Table 1). 
    \item Considering a task-specific scenario, we propose a strategy to jointly prune and fine-tune an RNN-T(RNN Transducer)-based ASR model, demonstrating that this approach performs better than pruning a pre-trained network first and then using it for ASR training.
    This improvement arises because gradually pruning the weights during ASR fine-tuning allows the network to dynamically recover and adjust to the aggressively pruned model weights, leading to better overall performance (Table~\ref{table:2}).
    \item We use state-of-the-art conformer-based cascaded architecture for model compression. This architecture is particularly suitable for streaming and non-streaming ASR systems.
\end{enumerate}
To the best of our knowledge, this is the first study to investigate the combined effect of KD loss and pruning techniques for model compression in conformer-based speech SSL models, addressing both streaming and non-streaming scenarios in ASR.
\section{Related Work}
\subsection{\textbf{Structured Pruning Using $l_0$ Regularization}}
\label{l0_method}
The $l_0$ regularization is a powerful technique for pruning the neural networks by directly penalizing the number of non-zero weights in the network.
% The key idea was proposed in~\cite{L0_reg_louizos2017learning}.
We provide the key idea of this method, while more detailed explanations can be found in~\cite{L0_reg_louizos2017learning}.
Consider a frozen teacher model $f^{tea}(\cdot)$.
Also consider the student model $f^{stu}(\bs{\theta})$ with learnable parameters $\bs{\theta}=\{\theta_j\}_{j=1}^{L}$. 
Each $\theta_j$ represents a group of parameters that can be pruned, such as convolution channels, attention heads, or linear projections, with a total of $L$ groups.
Consider a binary variable $z_j$ (called mask), and the pruned version of $\theta_j$ which is given by 
$\tilde \theta_j = \theta_j z_j$.
The masks $\bs{z} = \{z_j\}_{j=1}^{L}$, follow a probability distribution $q(\bs{z};\boldsymbol{\alpha})$ parameterized by $\boldsymbol{\alpha} =\{\alpha_j \}_{j=1}^{L}$. 
The parameter $\bs{\alpha}$ regulates the sparsity levels of the model weights.
The objective for $l_0$ regularized distillation is expressed as:
\begin{equation}
\label{equation:L0}
\min_{\substack{\bs{\theta}, \bs{\alpha}}}
E_{\bs{z} \sim q} \left[\frac{1}{N} \sum_{i=1}^{N} \mathcal{L}_{dist}\left(f^{tea} (\bs{x_i}),f^{stu} (\bs{x_i},\bs{\Tilde{{\theta}}}) \right) + \lambda||\bs{\Tilde{{\theta}}}||_0\right]
\end{equation}
where $\mathcal{L}_{dist}(\cdot)$ is the distillation loss which can be $l_1$, $l_2$, or cosine distances~\cite{distilhubert}~\cite{FitHuBERT},$\{\bs{x_i}\}_{i=1}^{N}$ are training samples and
$\lambda > 0$ is a regularization weight.
Directly optimizing the $l_0$ norm $\|\bs{\tilde\theta}\|_0$ in Equation~\eqref{equation:L0} is non-differentiable.
To address this, Louizos \emph{et al.} proposed a reparameterization trick that samples the binary variable $\bs{z}$ from the Hard Concrete distribution~\cite{L0_reg_louizos2017learning}:
\begin{align} 
&\bs{u} \sim \mathcal{U}(0, 1), \bs{s(\bs{\alpha})} = \mathrm{sigmoid} \left(\left(\log \frac{\bs{u}}{1-\bs{u}} + \log \bs{\alpha}\right)/\beta\right), \nonumber \\ 
&\bs{\bar{s}}(\bs{\alpha}) = (r - l) \cdot \bs{s}(\bs{\alpha}) + l, \ \bs{z} = \min(1, \max(0, \bar{\bs{s}}(\bs{\alpha})))\nonumber 
\end{align}
where $\beta$ is a constant that controls the sharpness of the distribution, $s(\bs{\alpha})$ and the constants $l < 0$ and $r > 0$ stretch it into the range $[l,~r]$, after which it's clamped to $[0, 1]$, making the loss differentiable.
The final loss function is formulated as follows:
\begin{equation}
\label{equation:lagrangian}
\begin{aligned}
%\[
\max_{\substack{\lambda_1, \lambda_2}} \min_{\substack{\bs{\theta}, \bs{\alpha}}} E_{\bs{z} \sim q} \left[\frac{1}{N} \sum_{i=1}^{N} \mathcal{L}_{dist}\left(f^{tea}(\bs{x_i}),f^{stu}(\bs{x_i},\bs{\Tilde{{\theta}}}) \right) \right] \\ + \lambda_1(\bs{s}(\bs{\alpha}) - t) + \lambda_2(\bs{s}(\bs{\alpha}) - t)^2 \
%\]
\end{aligned}
\end{equation}
where $\bs{s}(\bs{\alpha})$ denotes the percentage of pruned parameters (the current sparsity-level), $t$ denotes the desired sparsity, and $\lambda_1$ and $\lambda_2$ are trainable Lagrangian multipliers~\cite{L0_reg_louizos2017learning}.
\subsection{\textbf{Structured Pruning Using Low-rank Factorization}}\label{LRF_method}
Structured pruning via low-rank factorization~\cite{LRF_structured} imposes sparsity by decomposing a weight matrix \(\textbf{W} \in \mathbb{R}^{m \times n} \) into smaller sub-matrices  \(\textbf{A} \in \mathbb{R}^{m \times r} \) and \(\textbf{B} \in \mathbb{R}^{n \times r} \), where $r$ is the rank of factorized matrices and $r \leq min(m,n)$~\cite{lee2018deeptwist}. 
The factorization reduces the number of parameters from $mn$ (in $\bs{W}$) to $r(m + n)$ (in $\bs{A}$ and $\bs{B}$), it is desirable to have $r(m + n)\leq mn$.
The rank $r$ controls the degree of compression.
The redundant information in $\bs{A}$ and $\bs{B}$ is pruned using  the learnable masks $\bs{z}\in \mathbb{R}^{r \times 1}$ (Section~\ref{l0_method}) as follows:
\begin{align}
\bs{W} = (\bs{A}\odot \bs{z})\bs{B}^T = \sum_{k=1}^{r} z_k \bs{a_k} \bs{b_k}^T \in \mathbb{R}^{m \times n} \nonumber
\end{align}
 where $\odot$ denotes point-wise multiplication,  $\bs{a_k}$ and $\bs{b_k}$ represent the $k^{th}$ column of $\bs{A}$,  and $\bs{B}$, respectively.
In the implementation of this method, a linear layer $\bs{W}$ is decomposed into two linear layers $\bs{A}$ and $\bs{B}$, with the masks inserted between them.
In addition to the low-rank factorization, the objective function for learning the target sparsity of a model is identical to the $l_0$ regularizer, as specified in Equation~\eqref{equation:lagrangian}.
\par 
%\section{Proposed Approach}
% \begin{figure}[t!]
% \centering \includegraphics[width=0.6\textwidth,height=0.9\textheight,keepaspectratio]{figure.pdf}
% \caption{} \label{fig:duration}
%\end{figure}
\section{Model Compression}
% We outline the proposed framework for model compression in detail.
\subsection{\textbf{The Pre-trained Network}}\label{sec:pt_model}
We select a state-of-the-art model proposed in~\cite{narayanan2021cascaded} and aim to compress it. 
This architecture comprises two interconnected encoder blocks: \textbf{C}ausal (C) and \textbf{N}on-\textbf{C}ausal (NC) which are arranged in a cascaded fashion.
Each block consists of conformer layers. 
Each conformer layer consists of an attention layer~\cite{vaswani2017attention}, a convolution layer sandwiched between two feed-forward layers~\cite{gulati2020conformer}.
The cascaded design of these blocks enables the generation of ASR transcriptions in both streaming and non-streaming modes.
This capability is achieved by configuring one encoder to operate in causal mode for streaming output and the other in non-causal mode for non-streaming output.\par
To pre-train this architecture, we employed \textbf{BE}RT-based \textbf{S}peech \textbf{T}raining with \textbf{R}andom-projection \textbf{Q}uantizer (BEST-RQ)~\cite{best-RQ}, which uses \textbf{M}asked \textbf{L}anguage \textbf{M}odeling (MLM) objective utilizing the self-supervised learning approach~\cite{MLM_ASR}. \\
\subsection{\textbf{The ASR Model}}
We utilize an RNN-T-based architecture for our ASR system, where the encoder consists of the cascaded encoder architecture~\cite{narayanan2021cascaded}, the prediction network is a \textbf{L}ong \textbf{S}hort-\textbf{T}erm \textbf{M}emory (LSTM) layer~\cite{staudemeyer2019understanding} and the joint network is a linear projection layer that combines the outputs from both the encoder and the prediction network.
 \begin{figure}[t!]
 \centering     
 \includegraphics[width=0.48\textwidth]{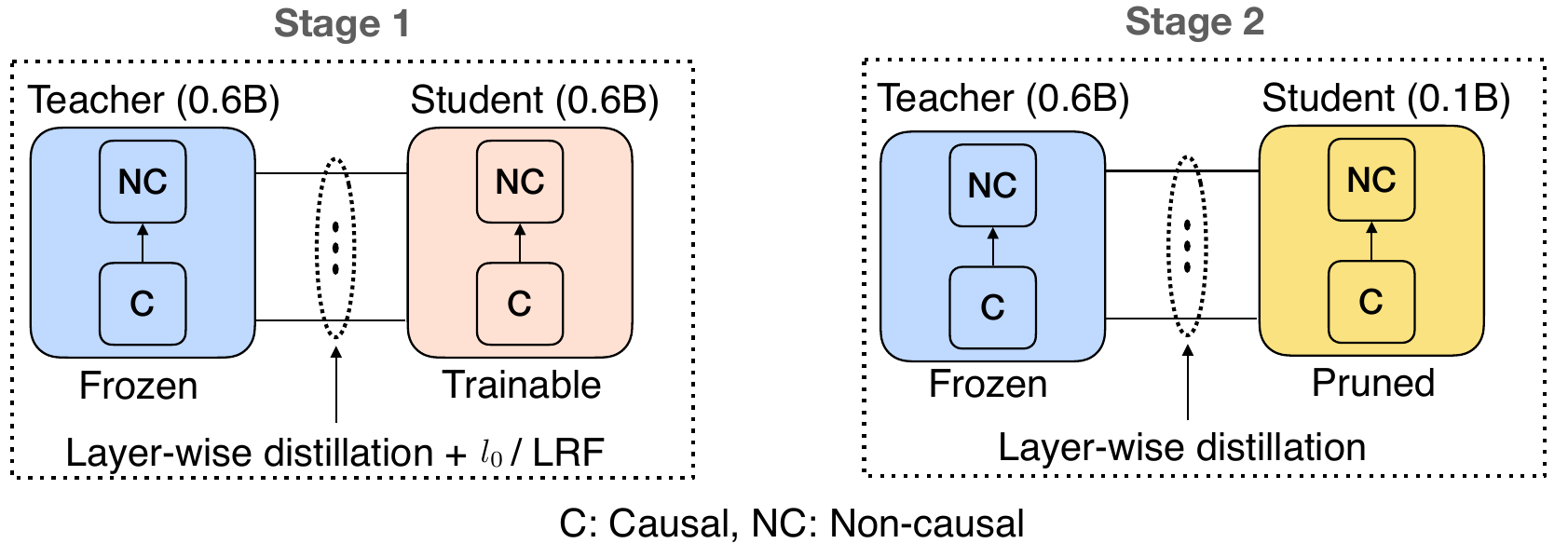}
     \caption{Pruning of the cascaded encoder architecture in the task-agnostic scenario. The compression is applied to both causal and non-causal blocks of the student model.}
  \label{fig:task_agnostic}
 \end{figure}
  \begin{figure}[t!]
 \centering     
 \includegraphics[width=0.43\textwidth]{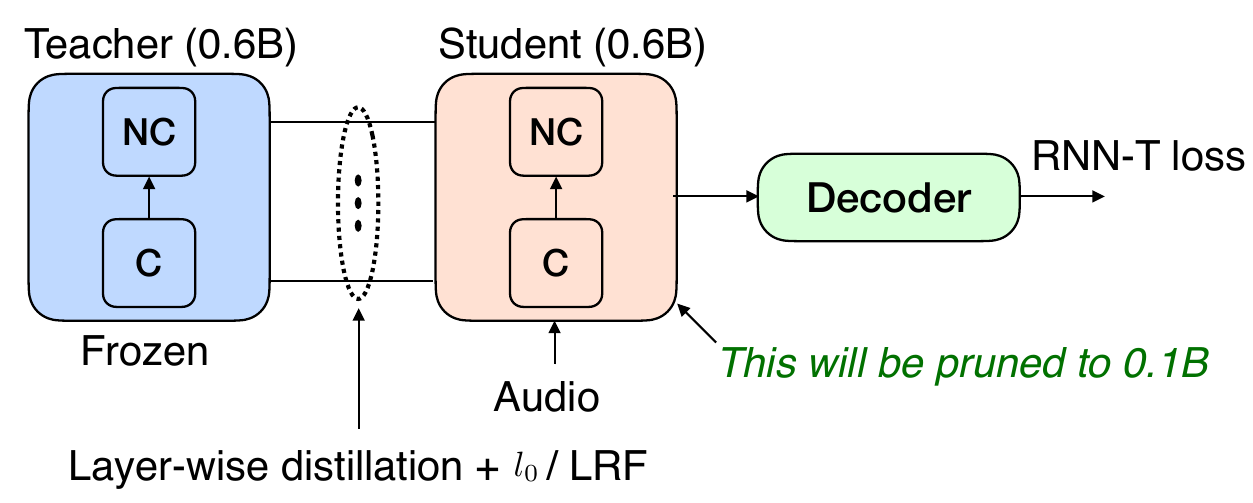}
     \caption{Joint pruning and fine-tuning of RNN-T-based ASR. The binary masks are learned during the process and subsequently used for pruning the student model.}
     \label{fig:task_specific}
 \end{figure}
\subsection{\textbf{Pruning of the Pre-trained Network}}{\label{sec:pt_prune}}
The network's parameters are predominantly dominated by dense layers (the linear projections), which account for $99.8\%$, while the remaining layers contribute only $0.2\%$.
Therefore, we concentrate our pruning efforts solely on the dense layers.\par 
Following~\cite{distilhubert, dphubert}, the pre-trained network undergoes a two-stage compression process, as depicted in Figure~\ref{fig:task_agnostic}. 
In Stage~$1$, a pre-trained network of size $0.6$ billion (B) parameters is used as the teacher, and the student model is initialized as its copy along with the masks $\bs{z}$ inserted into the conformer layers.
We apply the layer-wise distillation loss from Equation~\eqref{equation:lagrangian} which uses both $l_1$ and \textbf{C}osine \textbf{S}imilarity (CS) losses in an additive manner, each scaled by a factor of $0.5$.
We experimented by employing the layer-wise distillation loss for various combinations of the conformer layers and found optimal results by selecting every $5^{th}$ layer across all the $24$ layers, specifically layers $1, 5, 10, 15, 20$, and $24$.
When applying LRF, we decompose a dense layer into two separate dense layers, followed by the mask insertion between these two newly created layers. 
The masks $\bs{z}$ are learned during distillation.
The dense layer weights of the student model ($0.6$B) from Stage~$1$ are pruned by removing columns (rows) of weight matrices based on the learned masks, resulting in a smaller, pruned student model ($0.1$B parameters).
In Stage 2, we refine the distillation of the student model (0.1B) produced in Stage 1, employing only the layer-wise distillation component of the loss, as outlined in Equation~\eqref{equation:lagrangian} (see Figure~$1$).
This step is crucial for preserving key information that might be lost during pruning.
The distillation in Stage~$2$ is carried out for approximately half the steps used in Stage~$1$. 
Finally, the pruned student model ($0.1$B) is employed as an encoder in the RNN-T-based ASR.\par 
\subsection{\textbf{Joint Pruning and Fine-tuning in ASR}}
Figure~\ref{fig:task_specific} depicts the process of jointly pruning and fine-tuning an ASR system (task-specific scenario), with the key difference being the integration of the RNN-T loss~\cite{graves2012sequence} alongside the distillation loss ($l_1$ + CS) and the pruning methods (either $l_0$ or LRF). 
The masks $\bs{z}$ within the student model are learned simultaneously during ASR fine-tuning. 
Subsequently, the student model is pruned using the learned masks. \par 
In Section~\ref{sec:results}, we discuss the impact of combining KD either with $l_0$ or LRF on ASR performance.
In our experiments, KD is applied only between the outputs of the teacher and student model encoders.
\section{Experimental Setup}
\textbf{Dataset:}
We pre-trained the cascaded conformer model using the open-source VoxPopuli dataset~\cite{VoxPopuli}, which provides $400$K hours of unlabeled speech data across $23$ European languages. 
For fine-tuning, we utilized an in-house English dataset with $7$K hours of recordings from diverse environments, including clean, noisy (vehicle, room, cafe, factory), and telephone conditions. This dataset also includes MUSAN augmentation for various noise types and SpecAugment-based data augmentation~\cite{park2019specaugment}.\\
\textbf{Evaluation data:} We utilized a dataset of $10.2$ hours from $5$K utterances, consisting of $5.71$ hours of in-house recorded data to simulate real-world scenarios, and $4.49$ hours from $20$ distinct capsules (e.g., phone apps, calls, music) to account for variations such as dynamic speaker changes and noisy environments.
% -- bar plots ----
% \begin{figure*}[t]
%  \centering \includegraphics[width=1\textwidth,height=1\textheight,keepaspectratio]{ft_l0_lrf.png}
%  \vspace{-2mm}
%  \caption{The plot illustrates the sparse distribution of parameters remaining in Feed-Forward Block (FF1 \& FF2), Convolution Block (Conv), Attention Block (Self-Attention from 1–18th layers, and Relative Attention Block from 19–24th layers) at $83\%$ target sparsity.}
%   \label{fig:sparse}
% \end{figure*}
\\
\textbf{Implementation Details:}
% Based on~\cite{gulati2020conformer, zhang2023google},
We used a pre-trained model of size $0.6$B parameters as a teacher model and compressed it to a student model of size $0.1$B.  
The teacher model consists of 18 layers in the causal encoder and 6 layers in the non-causal encoder, each with a cell size of $1024$.
The ASR model used an LSTM layer with a dimension of $640$, a vocabulary size of $1024$, and employed byte-pair encoding for vocab generation~\cite{sennrich2015neural,singh2021comparative}.
Using LRF-based pruning, we initialized the rank $r=\frac{mn}{m + n}$ to ensure matching initial sizes for teacher and student encoders. 
For $l_0$-regularization based pruning (Section~\ref{l0_method}), we used $l = -0.1$ and $r = 1.1$, as adopted from~\cite{dphubert}.
We used $8$ NVIDIA V$100$ GPUs, compressing the model to $0.1$B parameters with an $83\%$ target sparsity in about $8$ GPU hours.
We employed the learning-rate scheduler with the initial learning rate of $1e^{-4}$ and warm-up of $5$K steps~\cite{vaswani2017attention}.
We used \textbf{W}ord \textbf{E}rror \textbf{R}ate (WER) as an objective measure for the ASR performance.
\section{Results}{\label{sec:results}}
Table~\ref{table:1} compares the performance of proposed pruning methods applied to a 0.6B pre-trained model (a task-agnostic scenario), followed by ASR fine-tuning. 
For baseline, we use a pre-trained model of size $0.1$B which was directly fine-tuned for ASR. 
The baseline model has $10$ and $6$ layers in causal and non-causal encoders, respectively, with a cell size of $512$. 
The row corresponding to $l_0$ uses a similar approach in DPHuBERT~\cite{dphubert}.
The $l_0$ and KD combination delivers the best non-streaming performance, showing a $6\%$ \textbf{R}elative \textbf{W}ord \textbf{E}rror \textbf{R}ate (RWER) improvement over the baseline. 
In contrast, LRF + KD achieves the highest performance for streaming ASR, with a $7.9\%$ improvement in RWER.
\par 
Table~\ref{table:2} presents a performance comparison of the proposed pruning methods in a task-specific scenario, where joint pruning and fine-tuning are applied to ASR.
We also include the performance results using the standard KD~\cite{panchapagesan2021efficient} approach for comparison.
Based on the observations from Table~\ref{table:1}, we only report the results for combining KD with either of the pruning methods. 
In Figure~\ref{fig:task_specific}, we examine two strategies for using the pre-trained network as a teacher model. 
The network can either be fine-tuned as an encoder in the RNN-T-based ASR before being used for distillation (PTFT-encoder in Table~\ref{table:2}), or it can be directly employed without fine-tuning (PT-encoder in Table~\ref{table:2}). 
The first strategy outperforms the second one by a small margin.
The $l_0$ + KD combination delivers the best non-streaming performance, with an $8.9\%$ RWER improvement over the baseline, while LRF + KD achieves a $13.4\%$ RWER improvement for streaming ASR.\par 
We observe that in both task-agnostic (Table~\ref{table:1}) and task-specific (Table~\ref{table:2}) settings, the combination of $l_0$ + KD consistently outperforms LRF + KD in non-streaming scenarios. 
However, LRF + KD shows superior performance in the streaming case. 
To understand this further, we examined the number of parameters retained after applying these pruning methods to both causal and non-causal encoder blocks, as reported in Table~\ref{table:3}. 
As mentioned in Section~\ref{sec:pt_model}, the non-streaming capability is attributed to the non-causal encoder, while streaming is due to the causal encoder. 
Table~\ref{table:3} shows that $l_0$ + KD retains more parameters in the non-causal encoder relative to LRF + KD, which aligns with the performance trends observed in Table~\ref{table:1} for the non-streaming case.
Conversely, LRF + KD retains more parameters in the causal encoder, consistent with the streaming performance trends observed in Table~\ref{table:1}.\par 
Tables~\ref{table:1} and~\ref{table:2} also present the computational performance in  \textbf{M}illion \textbf{F}loating \textbf{P}oint \textbf{O}perations (MFLOPs) of the pruned encoders using various pruning methods. 
The tables show that the proposed pruning methods result in no significant change in MFLOPs compared to the baseline.
\begin{table}[!htbp]
\captionsetup{width=\linewidth}  % Set caption width to match the table width
\caption{ASR performance of the pruned pre-trained model. We used a BEST-RQ pre-trained model ($0.6$B), pruned it to $0.1$B, and fine-tuned it for ASR.}
\centering
% \fontsize{10pt}{12pt}
\selectfont
\setlength{\tabcolsep}{1.2em}  % Adjust tabcolsep for better spacing
\begin{tabular}{lccc}
\hline
\textnormal{\bf{Method}} & \multicolumn{2}{c}{\bf{WER} (\%)} & \textnormal{\bf{MFLOPs}} \\ \cline{2-3}
& Non-streaming & Streaming \\ \hline
Baseline (0.1B) & 10.23 & 14.79 & 195.4 \\ \hline
$l_0$~\cite{dphubert} & 10.02 & 14.31 & 195.7 \\ 
LRF & 10.05 & 14.63 & 198.4 \\
$l_0$ + KD & \bf{9.57} & 13.91 & 195.5 \\
LRF + KD & 9.82 & \bf{13.62} & 198.7\\ \hline
\end{tabular}
\label{table:1}
\end{table}

% \begin{table}[!htbp]
% \captionsetup{width=0.47\textwidth}
% \caption{ASR performance of simultaneous pruning and fine-tuning. We used a 0.6B BEST-RQ pre-trained model, jointly pruned and fine-tuned it for ASR.}
% \centering
% \setlength{\tabcolsep}{0.48em}
% \begin{tabular}{lccc}
% \hline
% \textnormal{\bf{Method}} & \multicolumn{2}{c}{\bf{WER} (\%)} & {\bf{MFLOPS}} \\ \cline{2-3}
% & Non-streaming & Streaming &  \\ \hline
% Baseline (0.1B) & 10.23 & 14.79 & 192.8 \\ \hline
% KD (0.1B) & 12.56 & 16.48 & 192.8 \\ \hline
% $l_0$ + KD & \bf{9.31} & 13.39 & 188.8 \\ \hline
% LRF + KD   &  9.34 & \bf{12.84} & 197.9 \\ \hline
% \end{tabular}
% \label{table:2}
% \end{table}

\begin{table}[!htbp]
\caption{ASR performance of simultaneous pruning and fine-tuning. We used a BEST-RQ pre-trained model ($0.6$B), jointly pruned and fine-tuned it for ASR.}
\centering
% \fontsize{10pt}{12pt}
\selectfont
\setlength{\tabcolsep}{0.8em}  % Adjust tabcolsep for better spacing
\begin{tabular}{lccc}
\hline
\multicolumn{ 1}{l}{\bf{Method}} & \multicolumn{ 2}{c}{\bf{WER} (\%)}  & \textnormal{\bf{MFLOPs}} \\ \cline{ 2- 3}
\multicolumn{ 1}{c}{} & Non-streaming & Streaming \\ \hline
Baseline (0.1B)  & 10.23 & 14.79 & 195.4 \\
KD (0.1B)  & 12.56  & 16.48 & 195.4 \\ \hline
Teacher: PT-encoder &  &  \\ 
$l_0$  + KD  & 9.49 & 13.70 & 196.6 \\
LRF + KD & 9.60 & 13.35  & 197.9 \\ \hline
Teacher: PTFT-encoder &  &  \\ 
$l_0$  + KD  & \bf{9.31}  & 13.39 & 194.0 \\
LRF + KD  & 9.34  & \bf{12.8} & 199.4 \\ \hline
\end{tabular}
\label{table:2}
\end{table}

\begin{table}[!htbp]
\caption{Percentage of parameters retained after compression of the causal and non-causal encoders by various pruning methods, relative to the baseline model ($0.1$B).}
% \fontsize{10pt}{12pt}
\selectfont
\setlength{\tabcolsep}{0.85em}  % Adjust tabcolsep for better spacing
\begin{tabular}{lc|c|c|c}
\hline
\multicolumn{ 1}{c}{   } & \multicolumn{ 2}{c}{      Task-agnostic
 } & \multicolumn{ 2}{c}{    Task-specific
} \\ \cline{ 2- 5}
\multicolumn{ 1}{c}{} & L0 + KD & LRF + KD & L0 + KD & LRF + KD \\ \hline
Casual & 66\% & 68\% & 67\% & 73\% \\
Non-causal & $29$\% & 27\% & 28\% & 22\% \\ \hline
\end{tabular}
\label{table:3}
\end{table}
\section{Conclusions}
Firstly, we analyze the impact of combining knowledge distillation loss with either low-rank factorization or $l_0$ regularization-based pruning. 
Our results reveal that this combined approach consistently outperforms low-rank factorization or $l_0$ pruning alone.
Secondly, we propose and evaluate a novel strategy for task-specific ASR scenarios where pruning and fine-tuning are performed jointly. 
This method achieves notable performance gains, with the ability to adapt and recover from aggressive pruning during the fine-tuning phase being a key factor for these improvements.
Additionally, we explore the application of advanced conformer-based cascaded architectures for model compression, demonstrating their effectiveness for both streaming and non-streaming ASR systems.
This work is the first to combine knowledge distillation loss with pruning techniques in conformer-based speech SSL models, addressing both streaming and non-streaming scenarios.
Future work will examine the computational efficiency of deploying pruned models in real-time applications, focusing on the challenges associated with their practical implementation.
\bibliographystyle{IEEEbib}
\bibliography{strings,refs1}

\begin{thebibliography}{10}

\bibitem{wav2vec2}
Baevski \emph{et al.},
\newblock ``{W}av2{V}ec 2.0: A framework for self-supervised learning of speech
  representations,''
\newblock {\em Advances in neural information processing systems}, vol. 33, pp.
  12449--12460, 2020.

\bibitem{hubert}
Hsu \emph{et al.},
\newblock ``Hu{BERT}: Self-supervised speech representation learning by masked
  prediction of hidden units,''
\newblock {\em IEEE/ACM Transactions on Audio, Speech, and Language
  Processing}, vol. 29, pp. 3451--3460, 2021.

\bibitem{wavlm}
Chen \emph{et al.},
\newblock ``Wav{LM}: Large-scale self-supervised pre-training for full stack
  speech processing,''
\newblock {\em IEEE Journal of Selected Topics in Signal Processing}, vol. 16,
  no. 6, pp. 1505--1518, 2022.

\bibitem{data2vec}
Baevski \emph{et al.},
\newblock ``{D}ata2{V}ec: A general framework for self-supervised learning in
  speech, vision and language,''
\newblock in {\em Proc. of International Conference on Machine Learning}. PMLR,
  2022, pp. 1298--1312.

\bibitem{distilhubert}
Chang \emph{et al.},
\newblock ``Distil{H}u{BERT}: Speech representation learning by layer-wise
  distillation of hidden-unit {BERT},''
\newblock in {\em Proc. of IEEE International Conference on Acoustics, Speech
  and Signal Processing (ICASSP)}, 2022, pp. 7087--7091.

\bibitem{hinton2015distilling}
Hinton \emph{et al.},
\newblock ``Distilling the knowledge in a neural network,''
\newblock {\em arXiv preprint arXiv:1503.02531}, vol. 2, no. 7, 2015.

\bibitem{adiga2023compression}
Adiga \emph{et al.},
\newblock ``On the compression of shallow non-causal asr models using knowledge
  distillation and tied-and-reduced decoder for low-latency on-device speech
  recognition,''
\newblock {\em arXiv preprint arXiv:2312.09842}, 2023.

\bibitem{FitHuBERT}
Lee \emph{et al.},
\newblock ``Fit{H}u{BERT}: Going thinner and deeper for knowledge distillation
  of speech self-supervised models,''
\newblock in {\em Proc. of Interspeech}, 2022.

\bibitem{rathod2022multi}
Rathod \emph{et al.},
\newblock ``Multi-stage progressive compression of conformer transducer for
  on-device speech recognition.,''
\newblock in {\em proc. Interspeech}, 2022, pp. 1691--1695.

\bibitem{LRF_structured}
Ziheng Wang, Jeremy Wohlwend, and Tao Lei,
\newblock ``Structured pruning of large language models,''
\newblock {\em arXiv preprint arXiv:1910.04732}, 2019.

\bibitem{L0_reg_louizos2017learning}
Christos Louizos, Max Welling, and Diederik~P Kingma,
\newblock ``Learning sparse neural networks through $ l\_0 $ regularization,''
\newblock {\em arXiv preprint arXiv:1712.01312}, 2017.

\bibitem{lai2021parp}
Lai \emph{et al.},
\newblock ``{PARP}: Prune, {A}just and {R}e-{P}rune for self-supervised speech
  recognition,''
\newblock {\em Advances in Neural Information Processing Systems}, vol. 34, pp.
  21256--21272, 2021.

\bibitem{wang2024unstructured}
Haoyu Wang and Wei-Qiang Zhang,
\newblock ``Unstructured {P}runing and {L}ow {R}ank {F}actorisation of
  self-supervised pre-trained speech models,''
\newblock {\em IEEE Journal of Selected Topics in Signal Processing}, 2024.

\bibitem{dphubert}
Peng \emph{et al.},
\newblock ``D{PH}u{BERT}: Joint distillation and pruning of self-supervised
  speech models,''
\newblock {\em arXiv preprint arXiv:2305.17651}, 2023.

\bibitem{DeepversusWide}
Ashihara \emph{et al.},
\newblock ``Deep versus {W}ide: An analysis of student architectures for
  task-agnostic knowledge distillation of self-supervised speech models,''
\newblock in {\em Proc. of Interspeech}, 2022.

\bibitem{onceforall}
Han Cai, Chuang Gan, and Song Han,
\newblock ``Once for all: Train one network and specialize it for efficient
  deployment,''
\newblock {\em ArXiv}, vol. abs/1908.09791, 2019.

\bibitem{LightHuBERT}
Wang \emph{et al.},
\newblock ``Light{H}u{BERT}: Lightweight and configurable speech representation
  learning with once-for-all hidden-unit {BERT},''
\newblock in {\em Proc. of Interspeech}, 2022.

\bibitem{lee2018deeptwist}
Dongsoo Lee, Parichay Kapoor, and Byeongwook Kim,
\newblock ``Deep{T}wist: Learning model compression via occasional weight
  distortion,''
\newblock {\em arXiv preprint arXiv:1810.12823}, 2018.

\bibitem{narayanan2021cascaded}
Narayanan \emph{et al.},
\newblock ``Cascaded encoders for unifying streaming and non-streaming {ASR},''
\newblock in {\em Proc. of IEEE International Conference on Acoustics, Speech
  and Signal Processing (ICASSP)}. IEEE, 2021, pp. 5629--5633.

\bibitem{vaswani2017attention}
Vaswani \emph{et al.},
\newblock ``Attention is all you need,''
\newblock {\em Proc. NIPS}, vol. 30, 2017.

\bibitem{gulati2020conformer}
Gulati \emph{et al.},
\newblock ``Conformer: Convolution-augmented transformer for speech
  recognition,''
\newblock {\em arXiv preprint arXiv:2005.08100}, 2020.

\bibitem{best-RQ}
Chiu \emph{et al.},
\newblock ``Self-supervised learning with random-projection quantizer for
  speech recognition,''
\newblock in {\em proc. International Conference on Machine Learning}. PMLR,
  2022, pp. 3915--3924.

\bibitem{MLM_ASR}
Chung \emph{et al.},
\newblock ``W2v-{BERT}: Combining contrastive learning and masked language
  modeling for self-supervised speech pre-training,''
\newblock in {\em Proc. of ASRU}, 2021, pp. 244--250.

\bibitem{staudemeyer2019understanding}
Ralf~C Staudemeyer and Eric~Rothstein Morris,
\newblock ``Understanding {LSTM}--a tutorial into long short-term memory
  recurrent neural networks,''
\newblock {\em arXiv preprint arXiv:1909.09586}, 2019.

\bibitem{graves2012sequence}
Alex Graves,
\newblock ``Sequence transduction with recurrent neural networks,''
\newblock {\em arXiv preprint arXiv:1211.3711}, 2012.

\bibitem{VoxPopuli}
Wang \emph{et al.},
\newblock ``Vox{P}opuli: A large-scale multilingual speech corpus for
  representation learning, semi-supervised learning and interpretation,''
\newblock {\em arXiv preprint arXiv:2101.00390}, 2021.

\bibitem{park2019specaugment}
Park \emph{et al.},
\newblock ``Spec{A}ugment: A simple data augmentation method for automatic
  speech recognition,''
\newblock {\em arXiv preprint arXiv:1904.08779}, 2019.

\bibitem{sennrich2015neural}
Rico Sennrich, Barry Haddow, and Alexandra Birch,
\newblock ``Neural machine translation of rare words with subword units,''
\newblock {\em arXiv preprint arXiv:1508.07909}, 2015.

\bibitem{singh2021comparative}
Singh \emph{et al.},
\newblock ``Comparative study of different tokenization strategies for
  streaming end-to-end {ASR},''
\newblock in {\em Proc. of ASRU}, 2021, pp. 388--394.

\bibitem{panchapagesan2021efficient}
Panchapagesan \emph{et al.},
\newblock ``Efficient knowledge distillation for rnn-transducer models,''
\newblock in {\em Proc. ICASSP}, 2021, pp. 5639--5643.

\end{thebibliography}
\end{document}